\documentclass[twocolumn,letterpaper,floatfix]{revtex4}
\usepackage{graphicx,amsmath,amssymb,verbatim}

\begin{document}
\hyphenation{Ryd-berg}
\title{Guiding of Rydberg atoms in a high-gradient magnetic guide}

\author{M. Traxler$^1$}
\author{R. E. Sapiro$^1$}
\author{C. Hempel$^2$}
\author{K. Lundquist$^1$}
\author{E. P. Power$^1$}
\author{G. Raithel$^1$}

\affiliation{$^1$Department of Physics, University of Michigan, Ann Arbor, MI 48109, USA}
\affiliation{$^2$Institute for Quantum Optics and Quantum Information, Innsbruck A-6020, Austria}

\date{\today}
\begin{abstract}
We study the guiding of $^{87}$Rb 59D$_{5/2}$ Rydberg atoms in a linear, high-gradient, two-wire magnetic guide. Time delayed microwave ionization and ion detection are used to probe the Rydberg atom motion.   We observe guiding of Rydberg atoms over a period of $5$~ms following excitation.  The decay time of the guided atom signal is about five times that of the initial state.  We attribute the lifetime increase to an initial phase of $l$-changing collisions and thermally induced Rydberg-Rydberg transitions. Detailed simulations of Rydberg atom guiding reproduce most experimental observations and offer insight into the internal-state evolution.
\end{abstract}
\maketitle

There has been a recent surge of interest in cold Rydberg atoms in a linear trapping geometry.  Such systems present the possibility of creating one-dimensional spin chains by exciting atoms into high-lying Rydberg levels, which interact strongly due to their large dipole moments \cite{Mulken2007,Muller2008,Weimer2008}.  Rydberg crystals, which have been proposed in a frozen atomic gas using the Rydberg excitation blockade effect, may be an interesting application within a linear structure \cite{Pohl2010}.  Entangled Rydberg atoms prepared in a linear guiding geometry could act as a shuttle for quantum information \cite{Lukin2001,Saffman2010}.  A one-dimensional trap or guide for Rydberg atoms could be used to further these types of research.  Cold Rydberg atoms have been experimentally trapped using magnetic~\cite{Choi2005}, electrostatic~\cite{Hogan2008}, and light fields~\cite{Anderson2011}.  Conservative trapping of Rydberg atoms in magnetic atom guides has been theoretically investigated in \cite{Lesanovsky.2004a, Lesanovsky.2004c, Lesanovsky.2005a}.  Theoretical calculations also indicate the possibility of stationary Rydberg atoms confined in magnetic traps and  magnetoelectric traps~\cite{Hezel2006, Schmidt2007, Mayle2009}.  These systems would allow one to study Rydberg gases in a one-dimensional geometry.  The Rydberg-Rydberg interaction properties in such a system have been theoretically studied in~\cite{Mayle2007}.  In the present paper we report the first guiding of Rydberg atoms in a linear magnetic guide.

\begin{figure}[ht]
    \includegraphics[width=.38\textwidth]{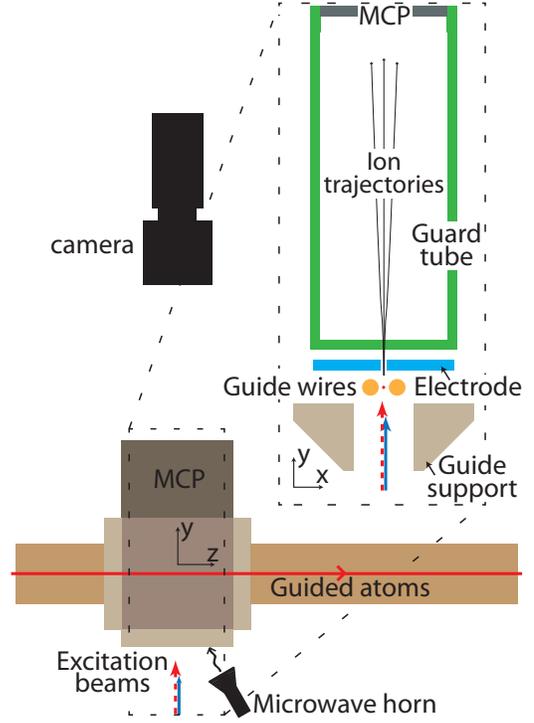}
    \caption{(Color online.)  Schematic of the excitation and detection region in the two wire magnetic guide. The guided atoms (small red dot) travel between the guide wires, where they are excited into Rydberg levels with two lower (780~nm, dashed red arrow) and one upper (480~nm, solid blue arrow) transition beams.  The Rydberg atoms are detected by microwave ionization.  With the guide support grounded and the guard tube at $-1.4$~kV, ions are directed upward to an MCP, where counts are recorded with both spatial and temporal resolution.  }
  \label{fig:guide}
\end{figure}

Two parallel wires carrying equal currents guide cold atoms in low magnetic field seeking states along a linear guiding channel located between the guide wires, where the magnetic field approaches zero.  Briefly, our 1.5~m long linear guide \cite{Olson.2006a,Mhaskar.2007a} operates with a magnetic field gradient of $\sim$1.5~kG/cm, which tightly confines a beam of $^{87}$Rb atoms (prepared in the $|F=1,m_F=-1 \rangle$ level of the $5$S$_{1/2}$ ground-state) in the guiding channel. The forward velocity of the guided atoms is adjusted to $\approx$1~m/s.  The ground state atoms have transverse and longitudinal temperatures of $T_{x,y}\approx400~\mu$K and $T_z\approx1~$mK, respectively.  An excitation and detection region for Rydberg atoms is located 85~cm down the guide, illustrated in Fig.~\ref{fig:guide}.   We use a three-step Rydberg atom excitation process.  A pulsed 780~nm beam (duration $10~\mu$s) pumps the atoms from 5S$_{1/2}$~$|F=1,m_F=-1 \rangle$ to $F=2$.  A second pulsed 780~nm beam (duration $5~\mu$s) subsequently drives the atoms on the cycling transition into 5P$_{3/2}$ $F'=3$.  The atoms are excited from the 5P$_{3/2}$ level to the $59$D$_{5/2}$ Rydberg level with a tunable, continuous 480~nm beam.  The electric field in the guiding region is zeroed by adjusting the potential of the electrode and the potential difference between the guide wires identified in Fig.~\ref{fig:guide}.

The Rydberg atoms are probed by ionization with a microwave pulse of $20~\mu$s duration and 18.5~GHz frequency.  The microwave pulse is applied at a variable delay time $t_{\rm d}$ and ionizes $>$$90\%$ of the Rydberg atoms. After ionization the free charges are decoupled from the microwaves.  With the electric field zeroed, about half of the ions drift downward and are lost.  The other half drift upward into the electric field generated by the guard tube, held at $-1.4$~kV, which accelerates them to a microchannel plate (MCP).  The MCP has an estimated ion detection efficiency of $30\%$, so the overall Rydberg atom detection efficiency is $\sim$15$\%$.  Each detected ion produces a countable pulse and a blip on a phosphor screen behind the MCP.

The temporal distribution of the MCP pulses is recorded using a multichannel scaler (MS). The MS traces presented in this paper are averages over up to 50000 experimental cycles.   The spatial
distribution is recorded with a CCD camera which images the blips on the MCP phosphor screen.  All images presented are an average of 30000 images with a background (30000 images without an
atomic beam) subtracted.

The ion distribution images have a transverse magnification of $M_{x}=-1.6$, where the negative sign indicates inversion in the $x$-direction due to an intermediate focal spot in the ion imaging \cite{Vaidya2010}.  This magnification is determined by moving the excitation beam by a known amount and measuring the resultant displacement in the image.  The ion images have a longitudinal magnification of $M_z=1$.  Because the atoms travel in the guide with a constant average velocity, longitudinal position in the images is linked with time.

\begin{figure}[ht]
    \includegraphics[width=.38\textwidth]{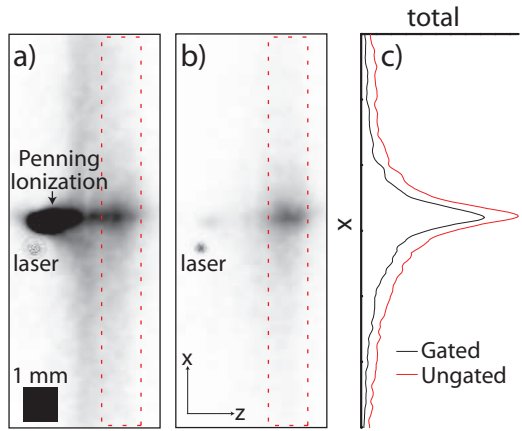}
    \caption{ MCP image with a delay time of $t_d=1.5~$ms, with (a) camera on throughout the
    experimental cycle and (b) camera gated over the microwave ionization signal.  The black square indicates 1 mm$^2$
    in the image plane.  The images exhibit a remnant of laser light penetrating through the MCP
    and phosphor screen. (c) Profiles of each, from the regions indicated by the dashed boxes. }
    \label{fig:gated}
\end{figure}

We detect an initial number of $\sim$30 ions per experimental cycle by integrating over the
microwave ionization signal for $t_{\rm d}=5~\mu$s (immediate detection).  Factoring in the
detection efficiency of 15\% and estimating that the excitation volume is a cylindrical region
of $50~\mu$m radius and $100~\mu$m height, the actual number of Rydberg atoms initially present
is $\sim$200 per experimental cycle and the corresponding density is of order 2$\times10^8$~cm$^{-3}$.
While this density is fairly low, it is sufficient to cause a small fraction of the atoms to
Penning ionize and to $l$-mix due to Rydberg-atom collisions. We note that our interaction times
are considerably longer than in other work \cite{Tanner2008}.

\begin{figure}[ht]
    \includegraphics[width=.5\textwidth]{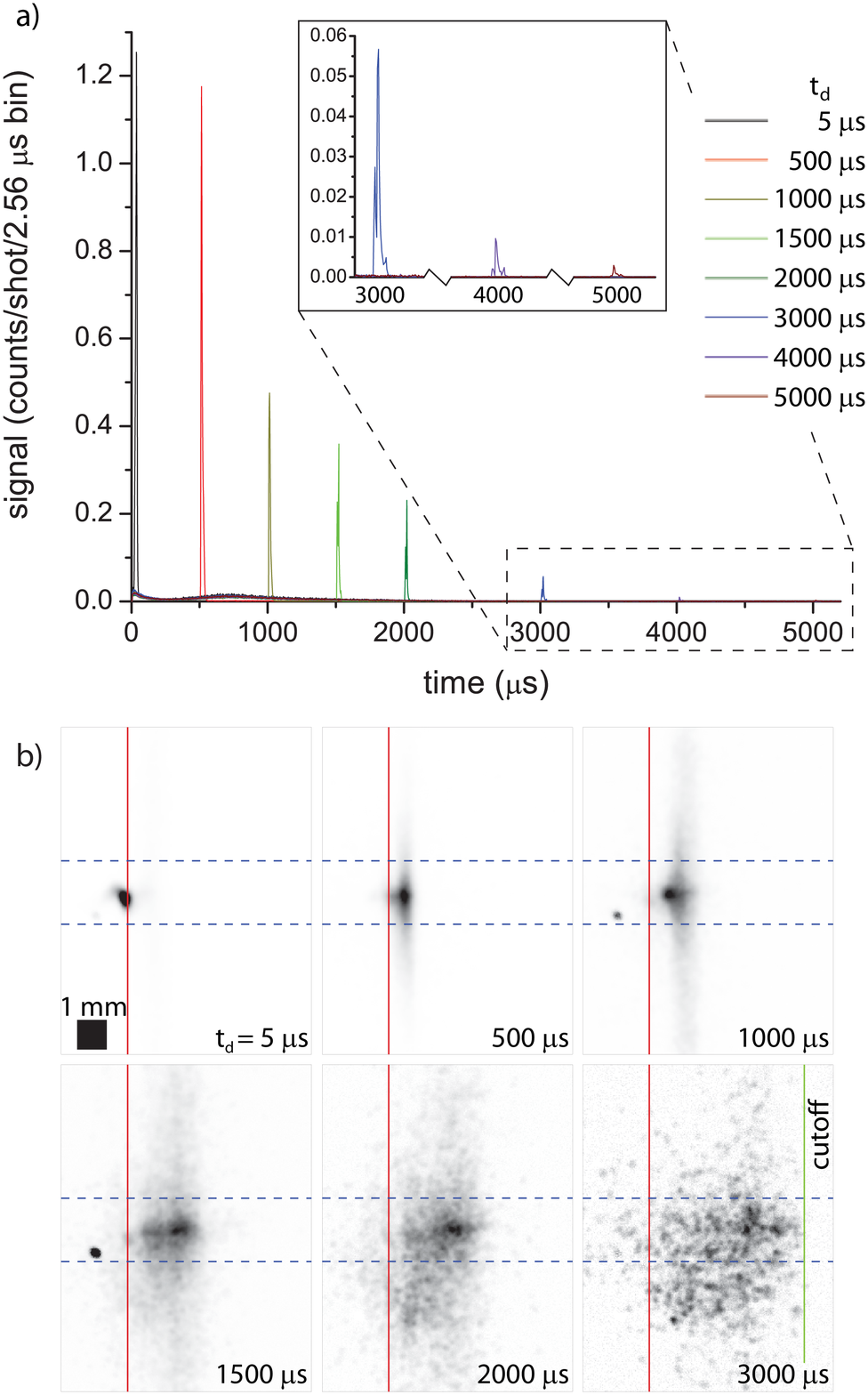}
    \caption{(Color online.) (a)  MS traces as a function of delay time $t_{\rm d}$. (b) Gated
    MCP images as a function of $t_{\rm d}$. The brightness scale of each picture is optimized
    separately for clarity.}
    \label{fig:long}
\end{figure}

Rydberg atoms that are initially excited into high magnetic field seeking states, or atoms that transition into such states, are expelled from the magnetic guide.  A fraction of these atoms enter the strong electric field region due to the guard tube, where they are field ionized. The resultant background ionization signal is fairly localized in time and does not significantly overlap with the microwave ionization signal in the MS traces (as shown in Fig.~\ref{fig:long}(a)).  However, the background signal is widely dispersed on the MCP and does spatially overlap with the Rydberg atom signal in the ion distribution images  (Fig.~\ref{fig:gated}(a)).  To eliminate the majority of the background signal in the images,  we usually gate the camera to selectively capture the microwave ionization component. In the gate timing, it is important to factor in the ion times of flight and the fluorescence decay of the phosphor screen.  Our ion times of flight range from $3$ to $25~\mu$s, peaking at $10~\mu$s.  These times are fairly long and reflect the small electric fields present in the atom  guiding region.  The fluorescence lifetime of the phosphor screen is $\approx50~\mu$s,
according  to manufacturer specifications.  We set the gate to range from just before the beginning of the microwave pulse to $150~\mu$s beyond its end.

In Fig.~\ref{fig:gated} we compare (a) an ungated and (b) a gated image, both taken at a microwave ionization delay time of $t_d=1.5~$ms.  The dominant
feature on the far left in the ungated picture is due to Penning ionization. Through ion counting measurements, we determine that $\sim$$7\%$ of the initially prepared Rydberg atoms Penning ionize.  The profiles in Fig.~\ref{fig:gated}(c), taken by integrating Figs.~\ref{fig:gated}(a) and (b) over the regions indicated by the dashed boxes, show that the gating removes about 17\% of the signal near the peak and more than 50\% in the wings.    Therefore, in the ungated image the signal in the peak is mostly due to microwave ionization, whereas a large fraction of the signal in the wings is from the background ion signal explained in the previous paragraph.  The gating procedure removes most of the background signal from the images.

We study the evolution of the Rydberg atoms in the guide by varying the delay time $t_{\rm d}$.  Figure~\ref{fig:long}(a) shows MS traces for different values of $t_{\rm d}$ ranging from 5~$\mu$s to 5~ms. Figure~\ref{fig:long}(b) shows a selection of gated images. We have verified that the signal in the images disappears when the microwave is turned off.  The integrated microwave ionization signal in the MS traces in Fig.~\ref{fig:long}(a) corresponds to the signal seen in the images and gives a quantitative measure for the number of Rydberg atoms present in the guide at time $t_{\rm d}$.

Based on the field gradient in the guide, we estimate that it takes a high magnetic field seeking Rydberg atom about 1~ms to become ejected from the guide. This estimate is confirmed in the simulations described below.  Hence, Rydberg atom signal observed after $\sim$1~ms is primarily attributed to guided Rydberg atoms.  As shown in the inset in Fig.~\ref{fig:long}(a) there is still microwave ionization signal at $t_{\rm d}=5~$ms, indicating that there is guiding of Rydberg atoms out to at least that time.  For $t_{\rm d}\geq3~$ms the image is partially clipped by the edge of a slit in the electrode noted in Fig.~\ref{fig:guide}, leading to a sharp cutoff in the ion images on the $+z$ side  (see $t_{\rm d}=3$~ms in Fig.~\ref{fig:long}(b)). The signal is only slightly cut off at $t_{\rm d}=3~$ms, but we miss detection of significant signal at $t_{\rm d}=4$~ms and most of it at $t_{\rm d}=5~$ms.

The red lines in Fig.~\ref{fig:long}(b) mark the $z$-location of immediate detection as a reference. The delayed guided Rydberg atom signals in Fig.~\ref{fig:long}(b) travel at an average speed of $1.2~$m/s, indicating that the average forward speed of those atoms does not change upon excitation to Rydberg states.  The velocity spread, given by the longitudinal extension of the ion signal divided by $t_{\rm d}$, remains constant, which implies that Rydberg-Rydberg collisions do not increase the temperature of the atomic sample beyond  $\sim$1~mK.

We note that the dashed lines in Fig.~\ref{fig:long}(b) represent the image location of the inner edges of the guide wires on the MCP. Signal observed outside of these boundaries should arise from atoms that are not in the guiding channel at the time when the microwave ionization is applied. Profiles of the images in the $x$-direction show that the
Rydberg-atom distribution transverse to the guiding channel maintains an approximately constant shape up to $t_{\rm d} = 1.5~$ms. At later delay times, the fraction of the signal outside the dashed lines in Fig.~\ref{fig:long}(b) increases.  This may indicate that at these later times an increasing fraction of un-guided atoms contributes to the microwave ionization signal. However, at all delay times we observe a dense core in the images that is due to guided Rydberg atoms in low magnetic field seeking states.

\begin{figure}[ht]
    \includegraphics[width=.5\textwidth]{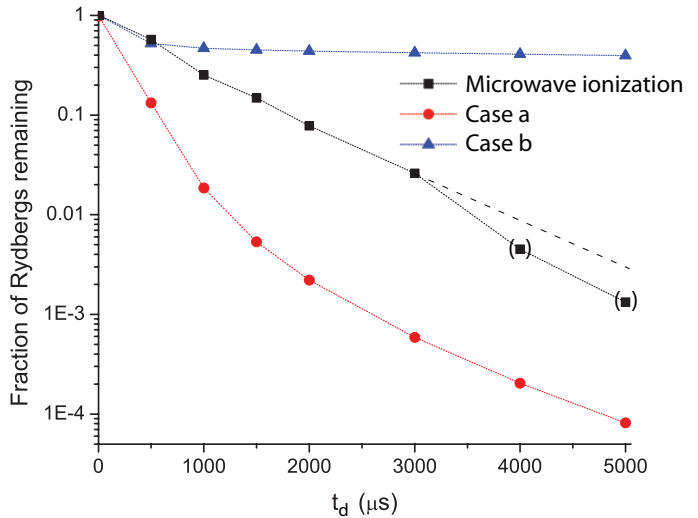}
    \caption{Fraction of Rydberg atoms remaining vs delay time, $t_{\rm d}$, from experiment (squares) and simulation with $l$-mixing (triangles) and without $l$-mixing (circles).  The parentheses indicate points where a significant portion of the signal is blocked from detection. The dashed line shows an extrapolation of the experimental curve based on the assumption that the Rydberg signal decay time does not change after $4~$ms, where the observable signal is partially obstructed.}
    \label{fig:log}
\end{figure}

Figure~\ref{fig:log} shows the fraction of microwave-ionized Rydberg atoms as a function of $t_{\rm d}$, with the experimental data plotted as squares.  The data consistently give a signal decay time of $\tau_{\rm{expt}}=700\pm150~\mu$s, excluding the data points at $t_{\rm d}=4$ and $5$~ms, where much of the signal is blocked by the edge of the slit.  The value of $\tau_{\rm{expt}}$ is significantly longer than the natural decay time of the $59$D$_{5/2}$ state, $\tau_{59\rm{D}5/2}=150~\mu$s, indicating a near-instantaneous increase in lifetime following excitation.

We compare the experimental data to a simulation in which the center-of-mass dynamics of the atoms are treated classically while the internal-state evolution is treated via a quantum Monte Carlo method. At each time step of the simulation, each Rydberg atom is in a well-defined state $\vert n,l,j,m_j \rangle$, with the quantization axis given by the direction of the magnetic field at the atom's location. Since the center-of-mass dynamics are several of orders of magnitude slower than the internal dynamics, we may assume that the internal states of the atoms adiabatically follow the changes in magnetic-field direction. Hence, the $m_j$-values and Land{\'e} g-factors of the atoms are well defined at all times, allowing for straightforward integration of the center-of-mass dynamics in the magnetic guiding field. Random numbers and appropriate transition probabilities are used to determine whether the atoms remain in their present states, photoionize, or transition into other Rydberg states, and also into which internal states the atoms transition if radiative transitions occur. All radiative transitions with rates $>1s^{-1}$ are accounted for.  A time step size of 5~$\mu$s is used, which is sufficiently short to accurately model the center-of-mass motion and the radiative transitions between the internal atomic states. The assumed radiation temperature is 300~K. The model accounts for guided atom loss due to thermal photoionization, decay toward the ground state, lifetime increases of the guided-atom population due to thermal transitions into states with higher $n$ and $l$, and  radiatively-driven probability flow between the guided domain ($m_j>0$) and the unguided domain ($m_j\leq0$).  The simulation does not account for Rydberg atom collisions with other particles (except in the initialization of the atomic sample).

Although the Rydberg atom samples we prepare in the experiment are not very dense, $\sim$$7\%$ of the Rydberg atoms Penning-ionize during the first few hundred microseconds, peaking at $10~\mu$s. The presence of Penning ionization suggests some degree of initial state-mixing of the remaining non-Penning-ionized atoms.  After the Penning ionization ceases, we expect collision-free evolution.  The simulations are therefore run for two types of initial conditions.  In case (a), the atoms are initially prepared in the 59D$_{5/2}$ level with randomly selected $m_j$ values.  Here, we model a collision-free situation in which the atoms evolve independently, solely due to radiative transitions between internal states and the magnetic force acting on the center-of-mass coordinates. In case (b), the atoms are initially prepared with an effective quantum number $n^*\approx 58$  and ($l$, $j$, $m_j$)-quantum numbers randomly selected with proper statistical weighting. There, we simulate the effects of an initial phase of strong $l$- and $m$-mixing due to Rydberg-atom collisions.

In Fig.~\ref{fig:log} we compare the experimentally observed fraction of remaining Rydberg atoms as a function of $t_{\rm d}$ with the simulation (case (a)--circles, case (b)--triangles). In simulation case (a), over the first $2~$ms the signal decay time gradually increases from about 200~$\mu$s to 1~ms, due to thermal transitions. In contrast, in the experiment the transition into longer lived Rydberg levels occurs on a near-instantaneous time scale.  The decay behavior of the experimental data and simulation case (a) match fairly well at $t_{\rm d}\gtrsim2~$ms; however, the experimental data consistently exceed the simulated ones by a factor of $\sim$50 in that time domain. We attribute this difference in behavior to the aforementioned Penning ionizing collisions, which apparently cause much faster, near-instantaneous state mixing and an associated rapid five-fold increase of the signal decay time.  The signal decay time in the simulation case (b), $\tau_{\rm{b}}=33\pm5~$ms, is about 30 times longer than the experimental value, indicating that the collision-induced $l$-mixing in the experiment falls far short of producing a shell-averaged sample.  We conclude that the experimental results lie between simulation cases (a)--no initial collision-induced mixing and (b)--maximal initial collision-induced $l$- and $m$-mixing.  Fig.~\ref{fig:log} further suggests that initial collisions do not populate any extremely long-lived high-$l$ states.

We have observed guiding of Rydberg atoms in a high-gradient linear magnetic guide for up to 5~ms.  To detect the atoms, we utilized a microwave ionization technique.   We found a guided Rydberg atom signal decay time nearly five times longer than the lifetime of the initial 59D$_{5/2}$ state.  In view of simulations, the increase in decay time is due to initial $l$-mixing collisions and thermal transitions.  In the future, one may realize much longer Rydberg atom guiding times by exciting long-lived circular-state Rydberg atoms in the guide using adiabatic state preparation techniques \cite{Nussenzweig1991}, which are quite compatible with our experimental apparatus.

\vspace{15 mm}

This work has been supported by the Army Research Office (50396-PH). MT acknowledges support by a NDSEG fellowship.

\vspace{15 mm}

\bibliographystyle{unsrt}

\end{document}